\documentclass[12pt]{article}
\usepackage{amsmath}
\usepackage{graphicx,psfrag,epsf}
\usepackage{enumerate}
\usepackage[round]{natbib}
\usepackage{url} 
\usepackage[dvipsnames]{xcolor}
\usepackage{amsmath,amssymb,amsfonts,amsthm}
\usepackage{float,graphicx, multirow, setspace}
\usepackage{subfiles}
\usepackage{placeins}
\usepackage{url}
\usepackage{hyperref}
\usepackage{xr, xr-hyper}
\usepackage{algpseudocode}
\usepackage{subfig}
\usepackage{algorithm}
\usepackage{makecell}
\usepackage[scr=boondox]{mathalfa}

\usepackage{array, multirow, bigdelim, makecell, booktabs}

 \renewcommand{\P}{{\mathbb P}}
 \newcommand{\ind}[1]{\mathbb{I}\left(#1\right)}

 \newcommand{\D}{{\mathscr{D} }}
 \newcommand{\bx}{{ \mathbf{x}}}

 \newcommand{\round}{{ \mathrm{round}}}
 \newcommand{\RMSE}{{ \mathrm{RMSE}}}
 \newcommand{\ESS}{{ \mathrm{ESS} }}
 \newcommand{\var}{{ \mathrm{var}}}
 \newcommand{\bias}{{ \mathrm{bias}}}
 \newcommand{\covg}{{ \mathrm{coverage}}}
 \newcommand{\width}{{ \mathrm{width}}}
 \newcommand{\CI}{{ \mathsf{CI}}}

\newcommand{\xgb}{{ \mathsf{XGBoost}}}

 \renewcommand{\wp}{{ \mathsf{WP}}}

\doublespace
\title{Exploring the Difficulty of Estimating Win Probability: A Simulation Study} 

\author{Ryan S. Brill\thanks{Graduate Group in Applied Mathematics and Computational Science, University of Pennsylvania. Correspondence to: ryguy123@sas.upenn.edu}, \ Ronald Yurko\thanks{Dept.~of Statistics and Data Science, Carnegie Mellon University}, \ and Abraham J. Wyner\thanks{Dept.~of Statistics and Data Science, The Wharton School, University of Pennsylvania}}

\begin{document}

\def\spacingset#1{\renewcommand{\baselinestretch}%
{#1}\small\normalsize} \spacingset{1}

\maketitle

\begin{abstract}
Estimating win probability is one of the classic modeling tasks of sports analytics.  Many widely used win probability estimators use machine learning to fit the relationship between a binary win/loss outcome variable and certain game-state variables. To illustrate just how difficult it is to accurately fit such a model from noisy and highly correlated observational data, in this paper we conduct a simulation study.  We create a simplified random walk version of football in which true win probability at each game-state is known, and we see how well a model recovers it. We find that the dependence structure of observational play-by-play data substantially inflates the bias and variance of estimators and lowers the effective sample size. Further, to achieve approximately valid marginal coverage, win probability confidence intervals need to be substantially wide. Concisely, these are high variance estimators subject to substantial uncertainty. Our findings are not unique to the particular application of estimating win probability; they are broadly applicable across sports analytics, as myriad other sports datasets are clustered into groups of observations that share the same outcome.

% Estimating win probability is one of the classic modeling tasks of sports analytics.  Many widely used win probability estimators use machine learning to fit the relationship between a binary win/loss outcome variable and certain game-state variables. To illustrate just how difficult it is to accurately fit such a model from noisy and highly correlated observational data, in this paper we conduct a simulation study.  We create a simplified random walk version of football in which true win probability at each game-state is known, and we see how well a model recovers it. We find that the dependence structure of observational play-by-play data substantially inflates the bias and variance of estimators and lowers the effective sample size. Thus, we want to quantify uncertainty in win probability estimates, but typical bootstrapped confidence intervals are too narrow and don't achieve nominal coverage. According to our novel method, the fractional bootstrap, confidence intervals need to be substantially wide to achieve adequate coverage. Our findings are not unique to the particular application of estimating win probability; they are broadly applicable across sports analytics, as myriad other sports datasets are clustered into groups of observations that share the same outcome.

\end{abstract}

\newpage
\spacingset{1.45} % DON'T change the spacing!

% %%%%%%%%%%%%%%%%%%%%%%%%%%%%%
% \documentclass[USenglish]{article}  
% \usepackage[utf8]{inputenc}  
% \usepackage[big,online]{dgruyter} %values: small,big | online,print,work
% \usepackage{lmodern} 
% \usepackage{microtype}
% \usepackage[round]{natbib}
% \bibliographystyle{apalike}
% \input{header}
% \begin{document}
% %%%%%%%%%%%%%%%%%%%%%%%%%%%%%

% % %%%%%%%%%%%%%%%%%%%%%%%%%%%%%
% % \documentclass[12pt]{article}
% % \usepackage{amsmath}
% % \usepackage{graphicx,psfrag,epsf}
% % \usepackage{enumerate}
% % \usepackage[round]{natbib}
% % \usepackage{url} 
% % \usepackage[dvipsnames]{xcolor}
% % \input{header}
% % \doublespace
% % \bibliographystyle{apalike}
% % % do not change margins - should be 1 inch all around.
% % \addtolength{\oddsidemargin}{-.5in}%
% % \addtolength{\evensidemargin}{-.5in}%
% % \addtolength{\textwidth}{1in}%
% % \addtolength{\textheight}{1.3in}%
% % \addtolength{\topmargin}{-.8in}%
% % \begin{document}
% % %%%%%%%%%%%%%%%%%%%%%%%%%%%%%

%%%%%%%%%%%%%%%%%%%%%%%%%%%%%%%%%%%%%%%%%%%%%%%%%%%%%%%%%%%%%%%%%%%%%%%%%%%%%%%
%%%%%%%%%%%%%%%%%%%%%%%%%%%%%%%%%%%%%%%%%%%%%%%%%%%%%%%%%%%%%%%%%%%%%%%%%%%%%%%
%%%%%%%%%%%%%%%%%%%%%%%%%%%%%%%%%%%%%%%%%%%%%%%%%%%%%%%%%%%%%%%%%%%%%%%%%%%%%%%
\section{Introduction}
\label{sec:intro}

Win probability ($\wp$) as a function of game-state is a canonical value function in sports analytics \citep{baumer2023big}.
In-game win probability is the crux of strategic decision making––make the decision that maximizes win probability––and is central to live betting on game outcomes.
Fourth-down decision making in American football is a prime example: choose between a conversion attempt, field goal attempt, and punt according to win probability \citep{brill2024analytics}.

Win probability is not a counting statistic or an observable quantity.
Rather, it is defined by a model that is estimated from data.
Estimating win probability is one of the classic modeling tasks of sports analytics.
Win probability estimates arise broadly from one of three classes of models: simple mathematical models, probabilistic state-space models, and statistical models. 
%%%%%%%%% 
Mathematical models are closed-form win probability functions with a simplified structure.
% Though they are easy to fit and use, they rely on unreasonable assumptions (e.g., normality)  and work for a limited set of game-state variables.
State-space models simplify a game into a series of transitions between game-states, and transition probabilities are propagated into win probability by simulating games.
% When implemented correctly, state-space models are sensible ways to estimate $\wp$, though they are difficult in practice (e.g., it is difficult to carefully encode the convoluted rules of a sport into a set of states and actions, and it is difficult to estimate transition probabilities).
Finally, statistical models are fit entirely from historical data––given the results of a set of observed plays, they fit the relationship between a binary win/loss outcome variable and certain game-state variables using regression or machine learning approaches.
For a more thorough review of various ways of estimating win probability, see Section~\ref{sec:estimatingWP} and \citet{baumer2023big}.

Notably, statistical $\wp$ models are widely used today by analysts of American football \citep{lockNettleton,BaldwinWP}.
For instance, they form the foundation of open-source fourth-down recommendations \citep{brill2024analytics}.
They are widely assumed to be reasonable and trustworthy because of the modeling flexibility of machine learning algorithms.
Hence, in this paper we focus on statistical win probability estimators.
The binary win/loss outcome variable, however, is noisy and features a strong dependence structure.
In particular, each play in the same game shares the same draw of the win/loss outcome.
Accordingly, \citet{brill2024analytics} found that these estimators are subject to substantial uncertainty and produce wide confidence intervals, even when fit from a large dataset featuring $229,635$ first-down plays across $4,101$ games and $16$ years.
We suspect statistical $\wp$ estimators have high variance, exacerbated by the dependence structure of observational play-by-play data.

% Notably, statistical $\wp$ models are widely used today by analysts of American football \citep{lockNettleton,BaldwinWP}.
% For instance, they form the foundation of open-source fourth-down recommendations \citep{brill2024analytics}.
% These models are popular due to the accessbility of rich publicly available data (e.g., play-by-play data from nflFastR \citet{nflFastR}) and off-the-shelf machine learning tools (e.g., $\xgb$ \citet{xgboost}). 
% They are widely assumed to be reasonable and trustworthy because of the modeling flexibility of machine learning algorithms. 
% For these reasons, in this paper we focus on statistical win probability estimators.

% The binary win/loss outcome variable, however, is noisy and features a strong dependence structure––each play in the same game shares the same draw of the win/loss outcome.
% Accordingly, \citet{brill2024analytics} found that these estimators are subject to substantial uncertainty and produce wide confidence intervals, even when fit from a large dataset featuring $229,635$ first-down plays across $4,101$ games and $16$ years.
% We suspect statistical $\wp$ estimators have high variance, exacerbated by the dependence structure of observational play-by-play data.

% \vspace{0.5in}

To illustrate just how difficult it is to accurately fit a statistical win probability model from noisy and highly correlated observational data, in this paper we conduct a simulation study.
We create a simplified random walk version of football in which true win probability at each game-state is known.
Then, we see how well a statistical model recovers true win probability.
% To our knowledge, this is the first exploration of the efficacy of win probability estimators via a simulation study.
%%%%%%% 
We find that the dependence structure of observational play-by-play data inflates both the bias and variance of these estimators.
We also calculate the effective sample size of the observational dataset.
Due to the dependence structure, we have half as much data as we think.
Specifically, a $\wp$ estimator fit from a correlated dataset with $4,101$ games has the same accuracy as one fit from a dataset of independent outcomes with half as many games.
Finally, we explore the efficacy of bootstrapped confidence intervals in quantifying uncertainty in $\wp$ estimates.
Naive bootstrapped confidence intervals do not achieve nominal marginal coverage.
% Hence, we introduce the fractional bootstrap, which can be tuned to produce approximately valid coverage.
Tuning the bootstrap to produce approximately valid coverage, we find that to cover true win probability $90\%$ of the time, confidence intervals need to be substantially wide (i.e., a mean width of $6.3\%$ $\wp$).
Each of these findings emphasize the difficulty of estimating win probability using machine learning.

Our study aligns thematically with existing research on the impact of clustered observations in sports analytics.
Other researchers have found similar conclusions, though in the context of injury or biomechanics data rather than the context of estimating win probability: a strong dependence structure inflates the bias and variance of estimators.
\citet{ClusteredDataInSportsResearch} found that ignoring a clustered dependence structure can lead to ``misleading conclusions'' and confidence intervals that are too narrow; analyzing clustered data requires ``an increased sample size when compared to those without clustering.''
\citet{southAfricaInjury} similarly found that naive statistical techniques that do not account for such dependence structures are ``inappropriate.''
\citet{clusterInjury} studied the impact of clustering on drawing inferences from injury-surveillance data. They conducted a simulation study similar in spirit to ours, simulating injury datasets ``using varying degrees of observation clustering,'' comparing ``inferences made using traditional techniques with those made after accounting for clustering.''
Not accounting for the dependency structure resulted in ``flawed inferences'' and biased estimates, with the degree of bias increasing as the strength of the clustering increased.
Finally, \citet{STAYNOR2019420} found that statistical analysis of clustered biomechanics data requires ``statistically appropriate'' models beyond naively applied regression models that result in ``erroneous parameter estimates...which have the potential to mislead future research and real-world applications.''

The remainder of this paper is organized as follows.
% In Section~\ref{sec:randomWalkFootball} we specify \textit{random walk football}, including the rules of the game, how we generate historical play-by-play datasets for that sport, and our statistical win probability model.
In Section~\ref{sec:randomWalkFootball} we specify the rules of \textit{random walk football}.
In Section~\ref{sec:pbpDataset} we discuss how we generate random walk football play-by-play datasets.
In Section~\ref{sec:estimatingWP} we detail various ways of estimating win probability, including the statistical $\wp$ estimator that we consider throughout this paper.
In Section~\ref{sec:biasVariance} we compute the bias-variance decomposition of a win probability estimator fit from various versions of observational datasets.
In Section~\ref{sec:ESS} we use this bias-variance decomposition to compute the effective sample size of a dataset that mimics the historical dataset of real American football plays.
In Section~\ref{sec:bootCovg} we show that naive bootstrapped win probability confidence intervals do not achieve nominal coverage and are too narrow.
In Section~\ref{sec:calibrateBoot} we introduce the fractional bootstrap, which we tune to produce wider confidence intervals that achieve adequate coverage in our simulation setting.
Finally, in Section~\ref{sec:discussion} we conclude and discuss ideas for future work.
% In particular, we look forward to further analyses of datasets in which groups of observations share the same outcome, which is prevalent across myriad sports datasets.    

%%%%%%%%%%%%%%%%%%%%%%%%%%%%%%%%%%%%%%%%%%%%%%%%%%%%%%%%%%%%%%%%%%%%%%%%%%%%%%%%%%
\section{The simulation study}
\label{sec:sim_study_wp}

%%%%%%%%%%%%%%%%%%%%%%%%%%%%%%%%%%%%%%%%%%%%%%%%%%%%%%%%%%%%%%%%%%%%%%%%%%%%%%%%%%
\subsection{Introducing random walk football}
\label{sec:randomWalkFootball}

We begin by describing the rules of random walk football.
Random walk football begins at midfield (yardline $L/2$, where $L$ is an even integer).
Each play, the ball moves left or right by one yardline with equal probability.
If the ball reaches the left end of the field (yardline $0$), team one scores a touchdown, worth $+1$ point.
If the ball reaches the right end of the field (yardline $L$), team two scores a touchdown, worth $-1$ point.
The ball resets to midfield after each touchdown.
After $T$ plays, the game ends.
If the game is still tied after $T$ plays, a fair coin is flipped to determine the winner.

% We include the formal mathematical specification of the game in Appendix~\ref{sec:sim_details}.
% We also explicitly compute true win probability as a function of time, field position, and score differential using dynamic programming in Appendix~\ref{sec:sim_details}.

%%%%%%%%%%%%%%%%%%%%%%%%%%%%%%%%%%%%%%%%%%%%%%%%%%%%%%%%%%%%
Formally, the outcome of the $t^{th}$ play of the $g^{th}$ game is 
\begin{equation}
\xi_{gt} \overset{iid}{\sim} \pm 1.
\end{equation}
The game starts at midfield, $X_{g0} = L/2$, and the game begins tied, $S_{g0} = 0$.
The field position at the start of play $t$ is 
\begin{equation}
X_{g,t+1} := 
\begin{cases} 
X_{gt} + \xi_{gt} & \text{ if } 0 < X_{gt} + \xi_{gt} < L \text{ (not a TD)} \\
L/2 & \text{ else,} 
\end{cases}
\end{equation}
and the score differential at the start of play $t$ is 
\begin{equation}
S_{g,t+1} := 
\begin{cases} 
S_{gt} + 1 & \text{ if } X_{gt} + \xi_{gt} = 0 \text{ (TD)} \\
S_{gt} - 1 & \text{ if } X_{gt} + \xi_{gt} = L \text{ (opp. TD)} \\
S_{gt}     & \text{ else.} 
\end{cases} 
\end{equation}
The binary win/loss response column is
\begin{equation}
\label{eqn:outcome_y}
y_{gt} \equiv y_{g,T+1} := 
\begin{cases}
1 & \text{ if } S_{g,T+1} > 0 \\
0 & \text{ if } S_{g,T+1} < 0 \\
\text{Bernoulli}(1/2) & \text{ else (overtime). }
\end{cases}
\end{equation}
% As in our dataset of real football plays, this response column is highly correlated––plays from the same game share the same draw of the winner of the game.
The true win probability 
\begin{equation}
\wp(t,x,s) := \P( S_{g,T+1} > 0 | X_{gt} = x, S_{gt} = s)
\end{equation}
of random walk football is computed explicitly using dynamic programming,
\begin{equation}
\wp(T+1,x,s) = 
\begin{cases}
    1 & \text{if } s > 0 \\
    1/2 & \text{if } s = 0 \\
    0 & \text{if } s < 0, \\
\end{cases} 
\end{equation}
and
\begin{equation}
\wp(t-1,x,s) =
\begin{cases}
    \frac{1}{2}\wp(t, \frac{L}{2}, s+1) + \frac{1}{2}\wp(t, x+1, s)  &\text{if } x=1 \\ %\text{ (about to score)} \\
    \frac{1}{2}\wp(t, x-1, s) + \frac{1}{2}\wp(t, \frac{L}{2}, s-1)  &\text{if } x=L-1 \\ %\text{ (opp. about to score)} \\
    \frac{1}{2}\wp(t, x-1, s) + \frac{1}{2}\wp(t, x+1, s)  & \text{else.} 
\end{cases}
\end{equation}
%%%%%%%%%%%%%%%%%%%%%%%%%%%%%%%%%%%%%%%%%%%%%%%%%%%%%%%%%%%%

%%%%%%%%%%%%%%%%%%%%%%%%%%%%%%%%%%%%%%%%%%%%%%%%%%%%%%%%%%%%%%%%%%%%%%%%%%%%%%%%%%
\subsection{Random walk football play-by-play data}\label{sec:pbpDataset}

A simulated observational dataset of random walk football plays consists of $G$ games, each with $T$ plays per game.
Such a dataset has the form
\begin{equation}
\D = \{ (t, X_{gt}, S_{gt}, y_{gt}) : t=1,...,T \text{ and } g = 1,...,G\}.
\label{eqn:observational_dataset}
\end{equation}
For each play of game $g$, we record the timestep $t$, the field position $X_{gt}$, the score differential $S_{gt}$, and a binary variable $y_{gt}$ indicating whether the team with possession wins the game.
The dependence structure of random walk football (and real football) play-by-play data manifests in the outcome vector $\{y_{gt}\}$: plays from the same game share the same draw of the winner of the game.
Formally, $y_{gt} \equiv y_{g,T+1}$ as in Equation~\eqref{eqn:outcome_y}.
Importantly, we also know the underlying true win probability $\wp_{gt}$ at each play, which we use to evaluate win probability estimates.

Throughout this study, we let $T=56$, the average number of first-down plays per game in the dataset of American football plays from \citet{brill2024analytics}.
We use $L=4$ yardlines so that the average number of plays between each score is similar to the average number of first-down plays in a game of American football.

We want to assess the impact of the dependence structure of the response variable of observational football data on the accuracy of a statistical win probability estimator.
To do so, we compare the accuracy of a $\wp$ estimator fit from datasets of varying degrees of dependence.
We introduce a parameter $K$ that controls the degree of dependence of the win/loss outcome variable in a generated dataset: we keep a random subsample of $K$ plays per generated game.
$K=1$––keep just $1$ randomly sampled play in each of the $G$ simulated games––reflects independence across all plays in the filtered dataset because each game is generated independently.
When $K=1$, the outcome variable $y_{gt}$ reflects an independent draw of the win/loss outcome of the game since each play is filtered from a separate game.
$K=T$––keep all $T$ plays in each of the $G$ simulated games––reflects full dependence within each game and is equivalent to the original dataset.
When $K=T$, the outcome variable $y_{gt}$ for each play $t$ in game $g$ reflects the same draw of the win/loss outcome of the game.
Integer values of $K$ between $1$ and $T$ reflect intermediate degrees of dependence because just $1 < K < T$ plays per game share the same draw of the response variable.
For concreteness, we visualize example datasets for $K=1$, $K=3$, and $K=T$ in Table~\ref{tab:example_K_datasets}.

%%%%%%%%%%%%%%%%
\begin{table}[hbt!]
    \centering
    \tiny %%% UNCOMMENT FOR ARXIV
    \begin{minipage}{0.32\linewidth}
        \centering
        \begin{tabular}{c c c cl}
            \toprule
            $x$ & $t$ & $s$ & $y$ \\
            \midrule
            1 & 1  & 0 & 1 & \hspace{-1em}\rdelim\}{0.4}{*}[ from game $1$] \\
            3 & 22  & 5 & 1 & \hspace{-1em}\rdelim\}{0.4}{*}[ from game $2$] \\
            3 & 43  & 1 & 1 & \hspace{-1em}\rdelim\}{0.4}{*}[ from game $3$] \\
            1 & 52  & 3 & 1 & \hspace{-1em}\rdelim\}{0.4}{*}[ from game $4$] \\
            \vdots & \vdots & \vdots & \vdots \\
            2 & 21 & -7 & 0 & \hspace{-1em}\rdelim\}{0.4}{*}[ from game $\zeta\cdot T - 1$] \\
            2 & 5  & 7 & 1 & \hspace{-1em}\rdelim\}{0.4}{*}[ from game $\zeta\cdot T$] \\
            \bottomrule
        \end{tabular}
        \caption{Visualizing a $K=1$ dataset.}
        \label{tab:sample_data_a}
    \end{minipage}
    \hfill
    \begin{minipage}{0.32\linewidth}
        \centering
        \begin{tabular}{c c c cl}
            \toprule
            $x$ & $t$ & $s$ & $y$ \\
            \midrule
            2 & 5  & 0 & 1 & \hspace{-1em}\rdelim\}{3}{*}[ from game $1$] \\
            1 & 16  & 1 & 1 \\
            3 & 37  & 3 & 1 \\
            %%%%%
            2 & 17  & 0 & 0 & \hspace{-1em}\rdelim\}{3}{*}[ from game $2$] \\
            2 & 22  & -1 & 0 \\
            1 & 54  & -4 & 0 \\
            \vdots & \vdots & \vdots & \vdots \\
            \multicolumn{4}{c}{\dots} \\
            1 & 3  & 0 & 1 & \hspace{-1em}\rdelim\}{3}{*}[ from game $\zeta\cdot T/3$] \\
            1 & 19  & 0 & 1 \\
            3 & 41  & 1 & 1 \\
            \bottomrule
        \end{tabular}
        \caption{Visualizing a $K=3$ dataset.}
        \label{tab:sample_data_b}
    \end{minipage}
    \hfill
    \begin{minipage}{0.32\linewidth}
        \centering
        \begin{tabular}{c c c cl}
            \toprule
            $x$ & $t$ & $s$ & $y$ \\
            \midrule
            2 & 1  & 0 & 1 & \hspace{-1em}\rdelim\}{8}{*}[ game $1$] \\
            3 & 2  & 0 & 1 \\
            2 & 3  & 0 & 1 \\
            1 & 4  & 0 & 1 \\
            \vdots & \vdots & \vdots & \vdots \\
            2 & 55 & 2 & 1 \\
            1 & 56 & 2 & 1 \\
            \multicolumn{4}{c}{\dots} \\
            2 & 1  & 0 & 0 & \hspace{-1em}\rdelim\}{8}{*}[ game $\zeta$] \\
            1 & 2  & 0 & 0 \\
            2 & 3  & 0 & 0 \\
            1 & 4  & 0 & 0 \\
            \vdots & \vdots & \vdots & \vdots \\
            3 & 55 & -5 & 0 \\
            2 & 56 & -5 & 0 \\
            \bottomrule
        \end{tabular}
        \caption{Visualizing a $K=T=56$ dataset.}
        \label{tab:sample_data_c}
    \end{minipage}
    \caption{
    Visualizing example generated datasets with varying degrees of dependence $K$.
    Each dataset has the same nominal sample size, $\zeta \cdot T$ plays (rows).
    The variables are: field position $x$, time $t$, score differential $s$, binary win/loss $y$.
    In the $K=1$ dataset (a), all plays are independent because they are generated from separate independent games.
    In the $K=3$ dataset (b), groups of 3 plays are generated from the same game.
    Each group of 3 plays shares the same draw of the response $y$ and plays from different games are independent.
    The $K=T$ dataset (c) includes all $T$ plays from each generated game.
    Within each game, all plays share the same draw of the response $y$, and plays from different games are independent.
    }
    \label{tab:example_K_datasets}
\end{table}
%%%%%%%%%%%%%%%%

Generating a dataset with $G$ games and $T=56$ plays per game, keeping just $K$ randomly sampled plays per game, has a nominal sample size (number of rows) of $G \cdot K$ plays.
We want to compare the performance of a win probability estimator fit from datasets of varying degrees of dependence $K$ that have the same nominal sample size.
To do so, given $K$, we generate $G = \round(\zeta \cdot T / K)$ games and keep $K$ plays per game.
This yields a dataset consisting of (approximately) $\zeta \cdot T$ plays, which is independent of $K$.
As $T=56$ throughout this paper, $\zeta$ parameterizes the sample size.

% Finally, continuing the tradition of \citet{BaldwinWP}, throughout this paper we estimate win probability using $\xgb$.
% The covariates are $\bx = (t,x,s)$, where $t$ denotes time, $x$ denotes field position, and $s$ denotes score differential.
% The outcome variable is binary win/loss $y$.
% We use half of the games from the training set as a validation set to tune $\xgb$ models.

%%%%%%%%%%%%%%%%%%%%%%%%%%%%%%%%%%%%%%%%%%%%%%%%%%%%%%%%%%%%%%%%%%%%%%%%%%%%%%%%
\subsection{Ways to estimate win probability}\label{sec:estimatingWP}

From a dataset of football plays, we want to estimate win probability.
%%%%%%%%%%%%%%%%%%%%%%%%%%%%%%%%%%%%%%%%%%%%%%%%%%%%%%%%%
Win probability is not a counting statistic or an observable quantity.
Rather, it is defined by a model that is estimated from data.
Estimating win probability is one of the classic modeling tasks of sports analytics.
Win probability estimates arise broadly from one of three classes of models: simple mathematical models, probabilistic state-space models, and statistical models.
We give an overview of these classes below.
For a more thorough review we refer to the reader to \citet{baumer2023big}.

% Win probability estimates arise broadly from one of three classes of models: simple mathematical models, probabilistic state-space models, and statistical models.
% %%%%%%%%%%%
% For example, \citet{stern94} uses a simple mathematical model to estimate win probability in basketball.
\citet{stern94}, for example, uses a simple mathematical model to estimate win probability in basketball.
Supposing possession-level score differential outcomes are independent and approximately Gaussian, he models the score differential process by a Brownian motion with drift $\mu$ points advantage for the home team and variance $\sigma^2$.
He uses probit regression to estimate $p(l,t)$, the probability the home team wins if they are leading by $l$ points after $t$ seconds of game time, $p(l,t) = \Phi\big( (l + (1-t)\mu) / \sqrt{(1-t)\sigma^2} \big)$.
The benefit of such mathematical models is their simplicity: they have closed-form solutions.
The drawback is they rely on unreasonable assumptions (e.g., normality), which fail in particular towards the end of a game and work for a limited set of game-state variables (e.g., just score differential and time).

State-space models simplify a game into a series of transitions between game-states.
Transition probabilities are estimated from play-level data and are then propagated into win probability by simulating games.
Win probability in baseball is commonly estimated using state-space models going back to \citet{doi:10.1080/01621459.1961.10480656}.
These models work well in baseball because the game consists of discrete events (i.e., the game is divided into 9 innings, each of which feature a sequence of individual pitcher-batter matchups) and there are just a few important game-state variables (e.g., base-state, outs, runs, and inning).
When implemented correctly, state-space models are sensible ways to estimate $\wp$. 
However, they are difficult in practice, as they require: a careful encoding of the convoluted rules of a sport into a set of states and the actions between those states, careful estimation of transition probabilities, and enough computing power to run enough simulated games to achieve desired granularity.
Each of these can be nontrivial depending on the complexity of the sport.

Finally, statistical models are fit entirely from historical data.
Given the results of a set of observed plays, statistical models fit the relationship between a binary win/loss outcome variable and certain game-state variables using regression or machine learning approaches.
Notably, these models are widely used today by analysts of American football \citep{lockNettleton,BaldwinWP}.
They form the foundation of open-source fourth-down recommendations \citep{brill2024analytics}.
These models are popular due to the accessibility of rich publicly available data (e.g., play-by-play data from nflFastR \citet{nflFastR}) and off-the-shelf machine learning tools (e.g., $\xgb$ \citet{xgboost}). 
They are widely assumed to be reasonable and trustworthy because of the modeling flexibility of machine learning algorithms. %, which allows them to capture interactions and nonlinearities given enough data.
For these reasons, in this paper we focus on statistical win probability estimators.
The binary win/loss outcome variable, however, is noisy and features a strong dependence structure--each play in the same game shares the same draw of the win/loss outcome.
Accordingly, \citet{brill2024analytics} found that these estimators are subject to substantial uncertainty and produce wide confidence intervals, even when fit from a large dataset featuring $229,635$ first-down plays across $4,101$ games and $16$ years.
We suspect statistical $\wp$ estimators have high variance, exacerbated by the dependence structure of observational play-by-play data.
%%%%%%%%%%%%%%%%%%%%%%%%%%%%%%%%%%%%%%%%%%%%%%%%%%%%%%%%%

Continuing the tradition of \citet{BaldwinWP}, throughout this paper we estimate win probability using $\xgb$.
The covariates are $\bx = (t,x,s)$, where $t$ denotes time, $x$ denotes field position, and $s$ denotes score differential.
The outcome variable is binary win/loss $y$.
We use half of the games from the training set as a validation set to tune $\xgb$ models.

%%%%%%%%%%%%%%%%%%%%%%%%%%%%%%%%%%%%%%%%%%%%%%%%%%%%%%%%%%%%%%%%%%%%%%%%%%%%%%%%
\subsection{Bias-variance decomposition}
\label{sec:biasVariance}

In this section, we analyze the bias-variance decomposition of an $\xgb$ win probability estimator.
We compare a $\wp$ estimator fit from datasets having the same nominal sample size $\zeta \cdot T$ but generated with varying degrees of dependence $K$.
Given a combination of data generating parameters, we generate $M=100$ training datasets.
We fit a win probability estimator from each dataset, $\{ \widehat{\wp}^{(m)} \}_{m=1}^{M}$.
We also generate $M=100$ out-of-sample testing datasets $\{\D_{\text{test}}^{(m)}\}_{m=1}^{M}$ using $(G=10,000, T=56, K=1)$.
Then, we calculate the squared bias of the $m^{th}$ estimator by
\begin{equation}
\bias^2_m = \frac{1}{|\D_{\text{test}}^{(m)}|} \sum_{\bx \in \D_{\text{test}}^{(m)}} \big(\wp(\bx) - \widehat{\wp}^{(m)}(\bx)\big)^2
\end{equation}
and the variance by
\begin{equation}
\var_m = \frac{1}{|\D_{\text{test}}^{(m)}|} \sum_{\bx \in \D_{\text{test}}^{(m)}} \big(\widehat{\wp}^{(m)}(\bx) - \frac{1}{M} \sum_{j=1}^{M} \widehat{\wp}^{(j)}(\bx)\big)^2.
\end{equation}
The root mean squared error is $\RMSE_m = \sqrt{\bias^2_m + \var_m}$.
We then calculate the average squared bias, variance, and $\RMSE$ across the $M$ simulations and their standard errors.

\begin{figure}[hbt!]
    \centering{}
    % {\includegraphics[width=\textwidth]{plots_sim_study/plot_bias_var_1A.png} }
    {\includegraphics[width=\textwidth]{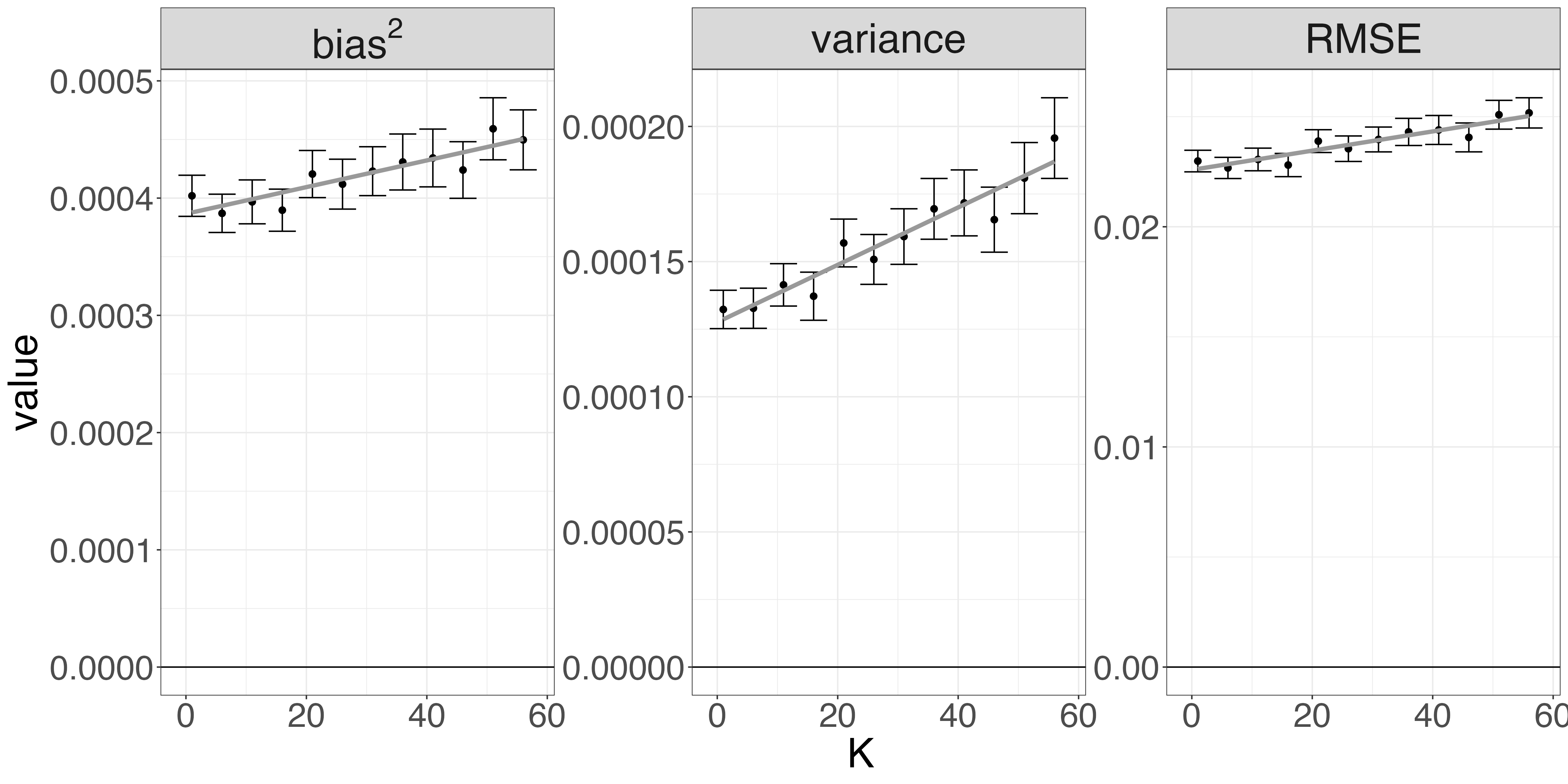} }
    \caption{
    Squared bias (left), variance (middle), and $\RMSE$ of a win probability estimator fit from a $(G = \round(\zeta \cdot T / K), \ T=56, \ K)$ dataset as a function of $K$, where $\zeta = 4,101$.
    The dots denote the average values across $M=100$ simulations and the bars denote plus/minus twice the standard errors.
    The gray line is the regression line.
    }
    \label{fig:plotBiasVarianceVaryK}
\end{figure}

First, let $\zeta = 4,101$ to mimic the dataset of real American football plays.
We compare the accuracy of a $\wp$ estimator fit from a $(G = \round(\zeta \cdot T / K), \ T=56, \ K)$ dataset as the degree of dependence $K$ varies.
The sample size in each of these datasets is (approximately) the same, $G \cdot K = \zeta \cdot T$.
We visualize this bias-variance decomposition as $K$ varies in Figure~\ref{fig:plotBiasVarianceVaryK}.
As the strength $K$ of the correlation increases, accuracy decreases linearly.
Fixing the number of observations in the dataset but increasing the degree of dependence across outcomes reduces model accuracy.
% We compute the explicit reduction in effective sample size in the next Section~\ref{sec:ESS}. 
Intuitively, this makes sense because the degree of dependence $K$ is inversely proportional to the amount of available independent data––there are (approximately) $\zeta \cdot T / K$ independent draws of the response variable.
Less independent data produces less accurate estimators.

%%%%%%%%%%%%%%%%%%
%%%%%%%%%%%%%%%%%%

Next, as a function of sample size $\zeta \cdot T$ (with $T = 56$), we compare the accuracy of a $\wp$ estimator fit from three types of datasets.
First, we consider a $(G = \zeta, \ K = T)$ dataset, which keeps each generated play per game.
This dataset mimics the historical dataset of real American football plays.
Then, we consider a $(G = \zeta, \ K = 1)$ dataset, which keeps just one randomly sampled play per game.
This dataset consists entirely of independent outcomes, and can be formed wholly from a $(G = \zeta, \ K = T)$ dataset, but its sample size is much smaller ($\zeta$ rather than $\zeta \cdot T$).
Finally, we consider a $(G = \zeta \cdot T, \ K = 1)$ dataset, which has the same sample size (number of rows) $\zeta \cdot T$ as the first dataset, but consists entirely of independent outcomes.

\begin{figure}[hbt!]
    \centering{}
    \subfloat[\centering \label{fig:plotBiasVarianceA}]{{\includegraphics[width=\textwidth]{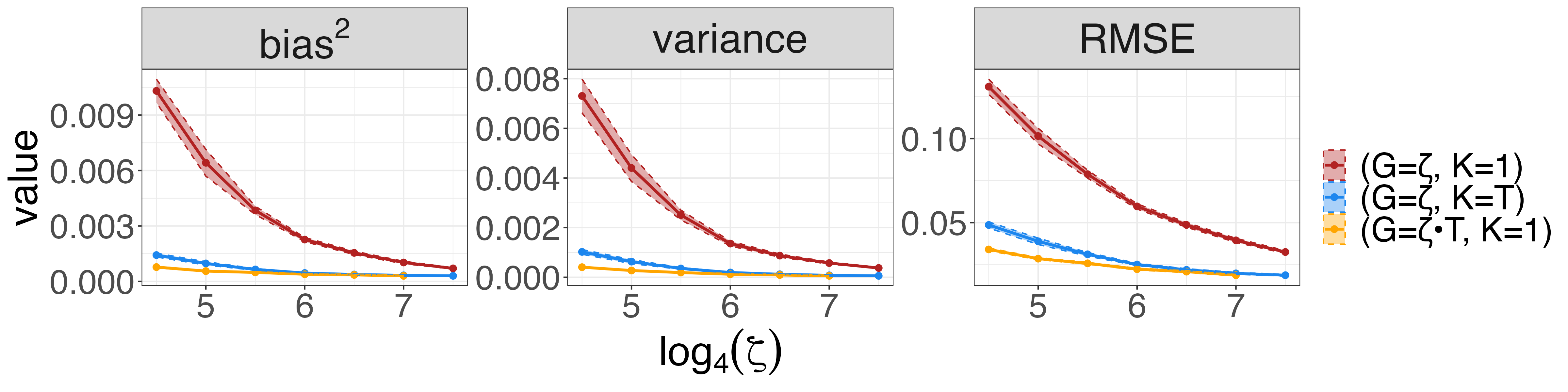} }}
    \\ 
    \subfloat[\centering \label{fig:plotBiasVarianceB}]{{\includegraphics[width=\textwidth]{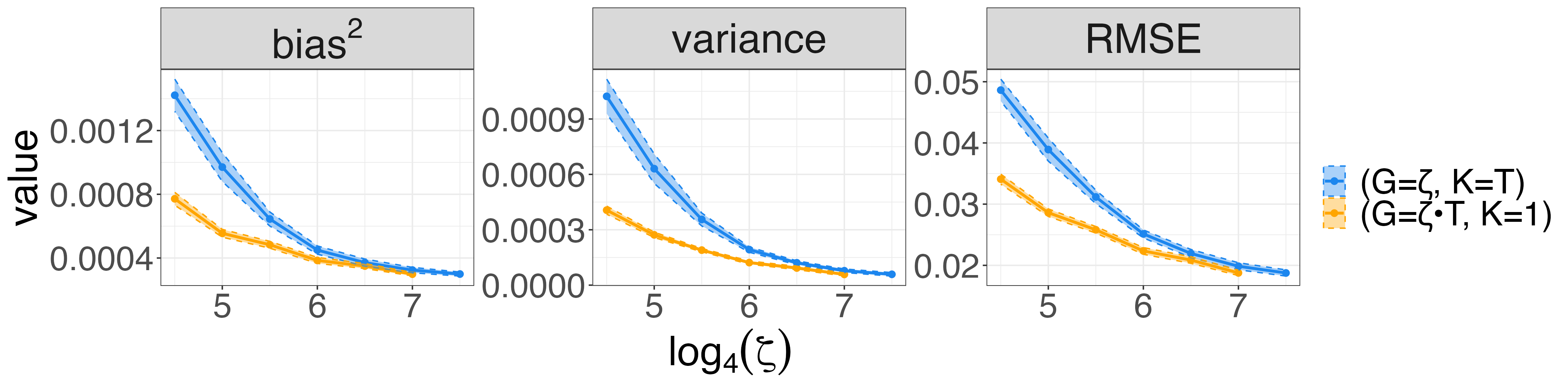} }}
    \caption{
    Squared bias (left), variance (middle), and $\RMSE$ (right) of a win probability estimator fit from three datasets as a function of $\zeta$.
    The $(G = \zeta, \ K = T)$ dataset (blue) involves keeping each generated play per game, which mimics the historical dataset of real American football plays.
    The $(G = \zeta, \ K = 1)$ dataset (red) is formed by keeping just 1 play per game.
    The $(G = \zeta \cdot T, \ K = 1)$ dataset (orange) has the same sample size (number of rows) $\zeta \cdot T$ as the first dataset but consists entirely of independent outcomes.
    }
    \label{fig:plotBiasVariance}
\end{figure}

In Figure~\ref{fig:plotBiasVariance} we visualize the bias-variance decomposition of a win probability estimator fit from these three types of datasets as a function of $\zeta$.
The $x$-axis is $\log_4(\zeta)$ because $4^6 = 4,096 \approx 4,101$ is the sample size of the historical dataset of American football plays.
We see that it is much better to use all plays per game rather than one independent play per game.
Despite the strong dependence structure, keeping all the plays elucidates information about the structure of the covariate space.
We also see that it would be much better if the plays had independent outcomes.
This suggests that the dependence structure reduces the effective sample size of the dataset.
We explore the extent of this reduction in the next section.

%%%%%%%%%%%%%%%%%%
%%%%%%%%%%%%%%%%%%

\begin{figure}[hbt!]
    \centering{}
    {\includegraphics[width=0.7\textwidth]{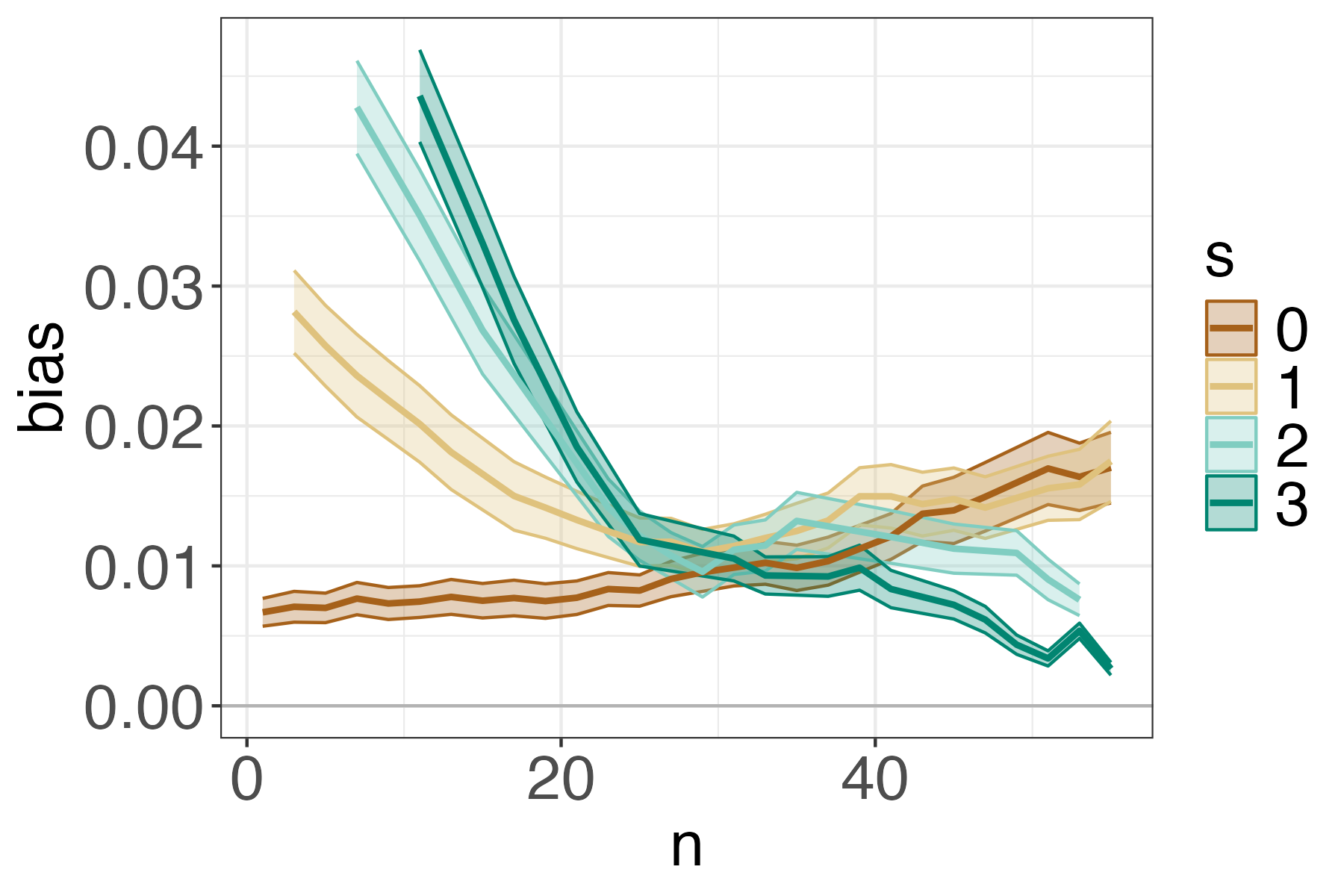} }
    \caption{
    Bias ($y$-axis) versus time $n$ ($x$-axis) and score differential $s$ (color) at midfield (field position $x=2$) of win probability estimated from a $(G = 4,101, \ T = 56, \ K = T)$ dataset.
    The lines denote the average values across $M=100$ simulations and the shaded regions denote plus-minus twice the standard errors.     
    }
    \label{fig:plotBiasInvestigation1}
\end{figure}

Interestingly, as shown in Figure~\ref{fig:plotBiasInvestigation1}, we see that the bias is much worse at the beginning of the game.
Game-states with larger score differentials in the early game occur less frequently than other game-states.
Also, intuitively it is easier to tie later game-states to the ultimate win/loss outcome, which is determined at the end of the game.

%%%%%%%%%%%%%%%%%%
%%%%%%%%%%%%%%%%%%

%%%%%%%%%%%%%%%%%%%%%%%%%%%%%%%%%%%%%%%%%%%%%%%%%%%%%%%%%%%%%%%%%%%%%%%%%%%%%%%%%%
\subsection{Effective sample size}
\label{sec:ESS}

As discussed in the previous section, we can calculate the accuracy of a win probability estimator fit from a $(G=\zeta, \ T=56, \ K)$ dataset, denoted $\RMSE(\zeta,K)$.
The sample size (number of rows) of such a dataset with $\zeta = 4,101$ and $K=T$, which mimics the historical dataset of American football plays, is $\zeta \cdot T$.
We saw that the dependence structure of this dataset reduces the accuracy of our estimator, but we would like to understand the extent of this reduction.
In particular, we are interested in the \textit{effective sample size} ($\ESS$) of that dataset.
The $\ESS$ is the sample size $\zeta' \cdot T$ of a $(G = \zeta' \cdot T, \ T = 56, \ K=1)$ dataset consisting of independent outcomes, which produces an estimator having the same accuracy as one fit from the original dataset.
For brevity, we refer to the sample size as $\zeta$ and the effective sample size as $\zeta'$, dropping the $T$ since we use $T=56$ throughout this study.

\begin{figure}[hbt!]
    \centering{}
    {\includegraphics[width=0.75\textwidth]{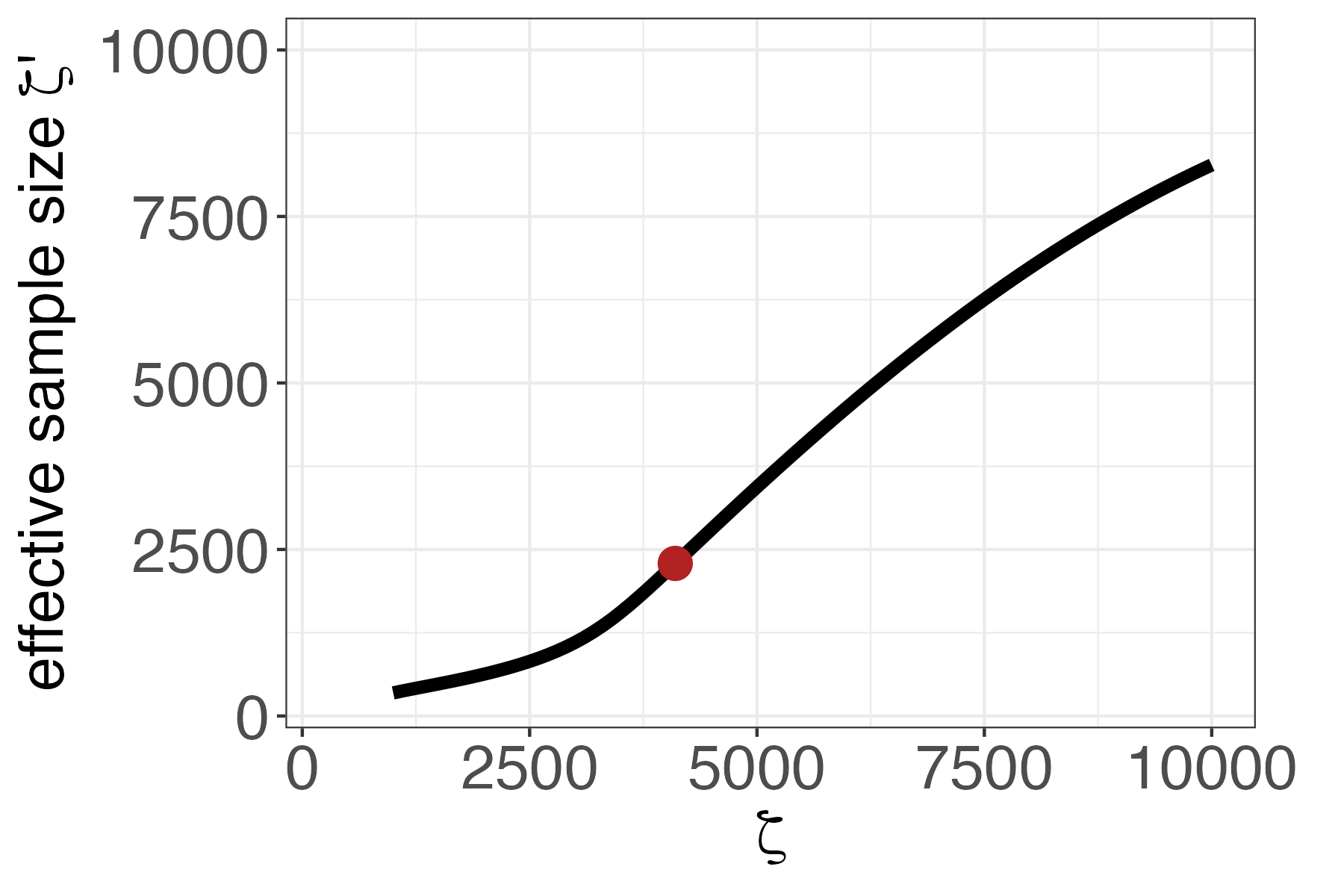} }
    \caption{
    The effective sample size $\zeta'$ of a $(G=\zeta, \ T = 56, \ K=T)$ dataset as a function of $\zeta$. 
    The red dot denotes $\zeta = 4,101$, the number of games in the historical dataset of real American football plays.
    }
    \label{fig:ESScurve}
\end{figure}

To estimate this effective sample size, we begin by fitting the $K=1$ and $K=T$ accuracy curves $\zeta \mapsto \RMSE(\zeta,K)$  from Figure~\ref{fig:plotBiasVarianceB}.
For each curve, we fit a biexponential model using nonlinear least squares.
Then, as a function of $\zeta$, the $\ESS$ is the value $\zeta'$ satisfying $\RMSE(\zeta',K=1) = \RMSE(\zeta,K=T)$.
In Figure~\ref{fig:ESScurve} we visualize the $\ESS$ $\zeta'$ as a function of $\zeta$.
The $\ESS$ of a $(\zeta = 4,101, K = T)$ dataset is $\zeta' = 2,291$, or $56\%$ of the nominal sample size.
This result is striking: we estimate that the historical dataset of American football plays (where $\zeta = 4,101$) consists of about half as much data as suggested by the number of plays.
In other words, we are effectively fitting win probability models from $8$ years, not $16$ years, worth of independent win/loss outcomes.
Real American football is exponentially more complex than random walk football.
Its game-state space is much larger, so we expect the $\ESS$ to be even smaller in real life.

%%%%%%%%%%%%%%%%%%
%%%%%%%%%%%%%%%%%%

If we halved the size of our $K=T$ dataset, fitting a win probability model from $\zeta = 2,050$ games ($8$ seasons), we estimate the effective sample size is $\zeta' = 645$, or just $31\%$ of the nominal sample size.
This mimics fitting a win probability model from just recent data.
If we doubled the size of our $K=T$ dataset, fitting a win probability model from $\zeta = 8,202$ games ($32$ seasons), we estimate the effective sample size is $\zeta' = 6,911$, or $84\%$ of the nominal sample size.

%%%%%%%%%%%%%%%%%%%%%%%%%%%%%%%%%%%%%%%%%%%%%%%%%%%%%%%%%%%%%%%%%%%%%%%%%%%%%%%%%%
\subsection{Coverage of bootstrapped confidence intervals}
\label{sec:bootCovg}

We have seen that machine learning win probability estimators fit from noisy and highly correlated observational data have high variance.
Due to the dependence structure of historical football data, the effective sample size is much smaller than the nominal sample size.
Therefore, we want to quantify uncertainty in win probability point estimates.
The point estimates alone may not be trustworthy.
The bootstrap is a natural choice to capture such uncertainty since it is non-parametric and does not make strong assumptions.
Hence, in this section we explore the efficacy of bootstrapped win probability confidence intervals.

We begin with the standard (i.i.d.) bootstrap, which assumes each row (play) of the dataset is independently drawn.
In the standard bootstrap, each of $B$ bootstrapped datasets are formed by re-sampling $G\cdot T$ plays uniformly with replacement (recall $G$ is the number of games, $T$ is the number of plays per game, and $G\cdot T$ is the total number of plays in a random walk football observational dataset).
The assumptions of the standard bootstrap do not apply to observational play-by-play data due to its dependence structure.
Hence, we also try the cluster bootstrap, in which each of $B$ bootstrapped datasets are formed by re-sampling $G$ games uniformly with replacement, keeping each observed play within each re-sampled game.
Finally, in the randomized cluster bootstrap, each of $B$ bootstrapped datasets are formed by re-sampling $G$ games uniformly with replacement, and within each game re-sampling $T$ plays uniformly with replacement.

Each type of bootstrap produces $B$ bootstrapped datasets $\{ \D_{\text{train}}^{(b)} \}_{b=1}^{B}$ from the training dataset $\D_{\text{train}}$.
We then fit a win probability model to each bootstrapped dataset using $\xgb$, $\{\widehat\wp_b\}_{b=1}^{B}$.
From these, we form a $90\%$ confidence interval for $\wp(\bx)$ at game-state $\bx$ by the $5^{th}$ and $95^{th}$ quantiles of $\{\widehat\wp_b(\bx)\}_{b=1}^{B}$.
Letting $B=101$ in this section, we form a $90\%$ confidence interval by $[\widehat\wp_{(6)}(\bx), \ \widehat\wp_{(96)}(\bx)]$.
To avoid substantially low coverage near the extremes ($\wp(\bx) \approx 0$ or $\wp(\bx) \approx 1$), we widen our confidence intervals when $\widehat{\wp}(\bx) < 0.025$ to have a lower bound of 0 and when $\widehat{\wp}(\bx) > 0.975$ to have an upper bound of 1.

We evaluate the efficacy of these intervals by their coverage and width.
For each type of bootstrap (standard bootstrap, cluster bootstrap, and randomized cluster bootstrap) and each simulation $m \in \{1,...,M=100\}$, we compute the pointwise marginal coverage of bootstrapped confidence intervals,
\begin{equation}
\label{eqn:sim_covg}
    \covg_m = \frac{1}{|\D_{\text{test}}^{(m)}|} \sum_{\bx \in \D_{\text{test}}^{(m)}} \ind{ \wp(\bx) \in \CI(\bx) }.
\end{equation}
This is the proportion of plays in $m^{th}$ held-out dataset whose true win probability lies inside the confidence interval.
We also compute the mean width,
\begin{equation}
\label{eqn:sim_len}
    \width_m =  \frac{1}{|\D_{\text{test}}^{(m)}|} \sum_{\bx \in \D_{\text{test}}^{(m)}}  \width(\CI(\bx)).
\end{equation}
For each play in the $m^{th}$ held-out dataset, we calculate the width of the confidence interval, and then take the average across all the plays.
We report the average and the standard error of these values $\{\covg_m\}_{m=1}^{M}$ and $\{\width_m\}_{m=1}^{M}$ in Table~\ref{table:simBootCIResults0}.
To mimic the historical dataset of American football plays, each simulated dataset here consists of $G=4,101$ games, $T=56$ plays per game, and $K=T$ plays per game that share the same outcome.

%%%%%%%%%%%%%%%%
\begin{table}[hbt!]
\centering
\begin{tabular}{ lll } \hline
 $90\%$ $\CI$ method & $\covg$ & $\width$ \\ \hline
 standard bootstrap  & $0.60 \pm 0.01$ & $0.027 \pm 0.0005$ \\
 cluster bootstrap  & $0.71 \pm 0.01$ & $0.036 \pm 0.0004$ \\
 randomized cluster bootstrap & $0.76 \pm 0.01$ & $0.042 \pm 0.0003$ \\ \hline
\end{tabular}
\caption{
Pointwise marginal coverage and mean width of nominally $90\%$ confidence intervals formed from each type of bootstrap with $B=101$ bootstrapped re-samples.
We report these values averaged over the $M=100$ simulations plus/minus twice their standard errors.
Each simulation uses $G=4,101$, $T=56$, and $K=T$.
}
\label{table:simBootCIResults0}
\end{table}
%%%%%%%%%%%%%%%%

Even in our simplified setting of random walk football, each of these bootstrapped win probability confidence intervals are undercovered.
Intuitively, naive bootstraps produce intervals that are too narrow because they involve resampling from observed data––which entails re-using observations without generating new ones––thus exploring a strictly smaller subspace of the ($\mathbf{x},y$) space than the true sampling distribution.
In other words, the bootstrapped resampling distribution is a rough approximation of the true sampling distribution.
The naive standard bootstrap in particular achieves dismally low marginal coverage.
Even the randomized cluster bootstrap that accounts for the dependence structure does not achieve high enough coverage.
We suspect this coverage issue would be even worse for real American football, which is exponentially more complex than random walk football.

%%%%%%%%%%%%%%%%%%%%%%%%%%%%%%%%%%%%%%%%%%%%%%%%%%%%%%%%%%%%%%%%%%%%%%%%%%%%%%%%%%
% \subsection{Calibrating bootstrapped confidence intervals}
\subsection{The fractional bootstrap}
\label{sec:calibrateBoot}

Naive bootstrapped confidence intervals are not wide enough.
A natural question arises: how wide do confidence intervals need to be so that nominally $90\%$ intervals actually achieve $90\%$ marginal coverage?
Hence, in this section we explore alternative forms of the bootstrap to increase coverage.

The traditional method of tuning non-parametric bootstrapped confidence intervals is to calibrate the bootstrapped quantiles \citep{efron_boot_ci}.
For instance, instead of using the $\alpha/2^{th}$ and $(1-\alpha/2)^{th}$ quantiles of $\{\widehat\wp_b(\bx)\}_{b=1}^{B}$ to form a $1-\alpha$ confidence interval, use the $\beta/2^{th}$ and $(1-\beta/2)^{th}$ quantiles for some $\beta < \alpha$.
In order for this traditional calibration method to work, $B$ would have to be much larger than $101$, likely an order of magnitude larger (e.g., $B=1001$).
We, however, prefer to use lower values of $B$ (e.g., closer to $101$) for several reasons.
It is much better to keep $B$ small for applications that require evaluating bootstrapped predictions in real time.
For example, a bootstrapped fourth-down decision recommendation from \citet{brill2024analytics} takes about 15 seconds when $B=101$ and about $2.5$ minutes when $B=1001$.
The former can be run before a fourth-down play begins and the latter takes far too long.
Additionally, storing $1001$ machine learning models is much more expensive than storing $101$ of them.
For these reasons, in this study we stray away from the traditional bootstrap calibration method.

% Instead, we introduce a novel method to calibrate bootstrapped confidence intervals, the \textit{fractional bootstrap}.
Instead, we introduce an alternative method to calibrate bootstrapped confidence intervals, the \textit{fractional bootstrap}.
It has the same time and storage complexity as traditional bootstrap methods.
Specifically, we introduce a parameter $\phi \in (0,1]$ denoting the fraction of data to be re-sampled in generating a bootstrapped dataset.
By re-sampling less data than in the original dataset, we widen bootstrapped confidence intervals and increase coverage.
In the fractional standard bootstrap, we re-sample $T \cdot G \cdot \phi$ plays (rows) uniformly with replacement.
In the fractional cluster bootstrap, we re-sample $G \cdot \phi$ games uniformly with replacement, keeping each observed play within each re-sampled game.
Finally, in the fractional randomized cluster bootstrap, we re-sample $G \cdot \phi$ games uniformly with replacement, and within each game re-sample $T$ plays uniformly with replacement. 

In Table~\ref{table:simBootCIResults_phi} we report the results of applying the randomized cluster bootstrap to our simulation study for various values of $\phi$.
To mimic the historical dataset of American football plays, each simulated dataset consists of $G=4,101$ games, $T=56$ plays per game, and $K=T$ plays per game that share the same outcome.
As expected, lowering $\phi$ widens the confidence intervals and increases marginal coverage.
In order to achieve $90\%$ marginal coverage, $\phi$ needs to be as small as $0.35$.
Those intervals have a mean width of $6.3\%$.
This result is striking: in our simplified setting of random walk football, win probability confidence intervals need to be extremely wide to achieve approximately valid coverage.
This exemplifies the difficulty of accurately estimating win probability by fitting a machine learning model from noisy and highly correlated football game outcomes.
These estimators are subject to large uncertainty.

%%%%%%%%%%%%%%%%
\begin{table}[hbt!]
\centering
\begin{tabular}{ lll } \hline
 $\phi$ & $\covg$ & $\width$ \\ \hline
  $1$ & $0.76 \pm 0.01$ & $0.042 \pm 0.0003$ \\ 
  $0.75$ & $0.80 \pm 0.01$ & $0.047 \pm 0.0003$ \\
  $0.5$ & $0.85 \pm 0.01$ & $0.055 \pm 0.0004$ \\
  $0.35$ & $0.90 \pm 0.01$ & $0.063 \pm 0.0004$ \\ \hline
\end{tabular}
\caption{
Pointwise marginal coverage and mean width of nominally $90\%$ confidence intervals formed from the $\phi$-fractional randomized cluster bootstrap with $B=101$ bootstrapped re-samples for various values of $\phi$.
We report these values averaged over the $M=100$ simulations plus/minus twice their standard errors.
Each simulation uses $T=56$, $K=56$, and $G=4,101$.
}
\label{table:simBootCIResults_phi}
\end{table}
%%%%%%%%%%%%%%%%

\begin{figure}[hbt!]
    \centering{}
    {\includegraphics[width=\textwidth]{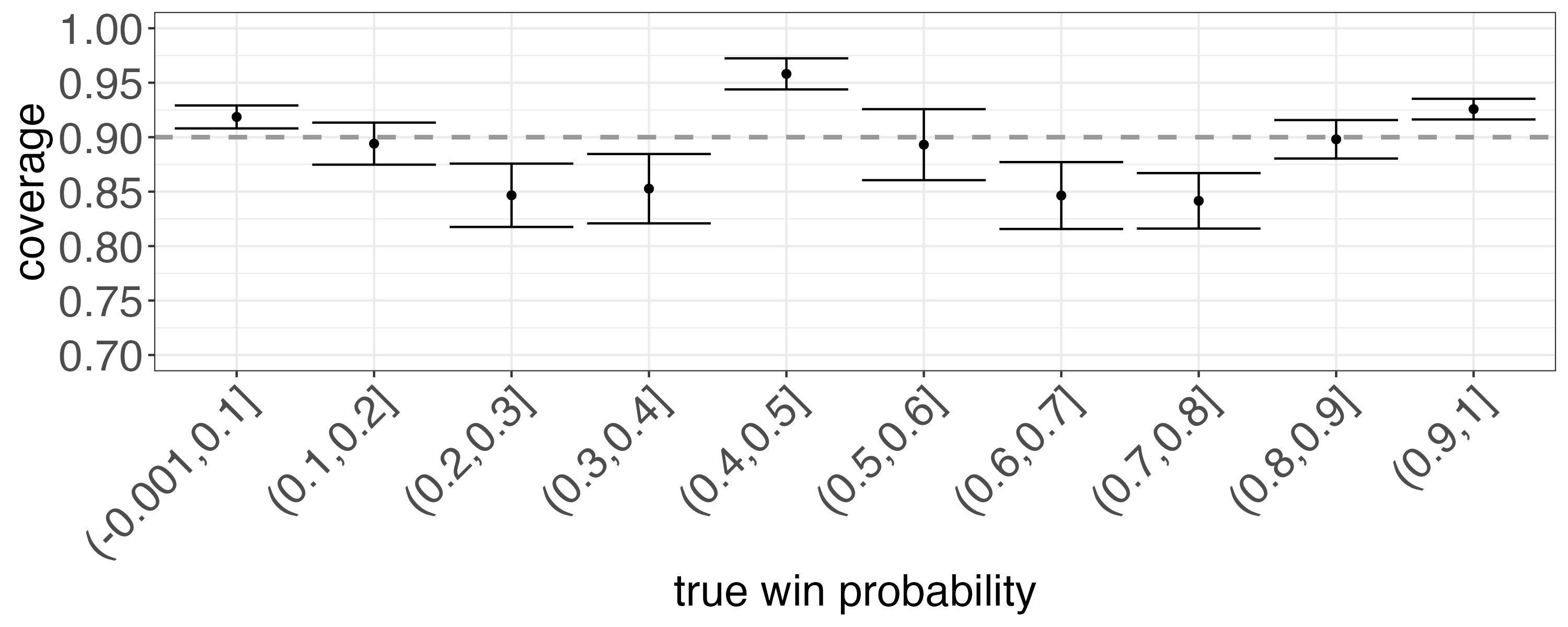} }
    \caption{
    Coverage of $90\%$ bootstrapped confidence intervals, via the fractional randomized cluster bootstrap for $\phi=0.35$, across the space of game-states $\bx$ binned by $\wp(\bx)$.
    }
    \label{fig:simBinnedCovg}
\end{figure}

Marginal coverage is a sufficient condition for confidence intervals to be ``good,'' but it is not a necessary requirement for decision making.
Even with $90\%$ marginal coverage, it could be that $\CI(\bx)$ always covers $\wp(\bx)$ for $90\%$ of game-states $\bx$ and never covers for the other $10\%$.
It may be disastrous to make decisions at the game-states for which intervals never cover.
To check that these intervals achieve reasonable coverage across the space of game-states, we bin game-states $\bx$ by their true win probability $\wp(\bx)$ and consider coverage in each bin.
In Figure~\ref{fig:simBinnedCovg} we visualize coverage and its standard error across such bins.
For bins near the middle ($\wp \approx 0.5$) or the extremes ($\wp \approx 0$ and $\wp \approx 1$), coverage is high enough.
For other bins  ($\wp \approx 0.3$ and $\wp \approx 0.7$), the intervals remain undercovered, albeit slightly (coverage hovers around $85\%$).
The game is still competitive in those regions and strategic decisions matter.
In future work, we recommend exploring more refined confidence intervals that achieve higher conditional coverage.

%%%%%%%%%%%%%%%%%%
%%%%%%%%%%%%%%%%%%

\vspace{0.5in}

In practice, it is impossible to tune the bootstrap at all, using either the traditional method from \citet{efron_boot_ci} or the fractional bootstrap.
This is because we need to know true win probability in order to tune the $\beta$ or $\phi$ values in these alternative bootstraps to achieve adequate coverage.
While we know true win probability in our simulation setting, it is an unobservable quantity in real life.
Coverage is not calculable for real data because the win/loss outcome is either $0$ or $1$ and win probability estimates lie in $(0,1)$.

This severe limitation of the tuned bootstrap, nonetheless, does not render its development in this paper worthless––it helped us further illustrate that win probability estimates are subject to substantial uncertainty.
Furthermore, though imperfect and difficult, we propose a way to tune the bootstrap with real data as follows.
Succinctly, we suggest tuning the fractional bootstrap using a hyper-realistic generative win probability model.
To do so, first, given the real historical play-by-play dataset, fit as realistic and granular a probabilistic state-space win probability model as possible.
As discussed in Section~\ref{sec:estimatingWP}, this entails fitting a play-level transition probability model, which is then propagated into win probability by simulating games.
Then, generate $M$ synthetic play-by-play datasets from the simulator, apply the $\phi$-fractional randomized cluster bootstrap for various values of $\phi$ to each of them, and select the value of $\phi$ that achieves desired marginal coverage.

As stressed in Section~\ref{sec:estimatingWP}, fitting a hyper-realistic football simulator is a delicate and extremely difficult task.
This is because it requires a careful encoding of the convoluted rules of football into a set of states and the actions between those states and careful estimation of transition probabilities.
Those we know who have created such simulators––including professional sports bettors, football analysts, and hedge fund analysts––do not make them publicly available because they are proprietary and because they use them to make money on the betting markets.
One contact said it took him eight months to construct such a simulator.

%%%%%%%%%%%%%%%%%%%%%%%%%%%%%%%%%%%%%%%%%%%%%%%%%%%%%%%%%%%%%%%%%%%%%%%%%%%%%%%%%%
\section{Discussion}\label{sec:discussion}

Statistical win probability estimators are widely used across sports analytics.
For instance, they form the foundation of open source fourth-down recommendations.
Here, we use a simulation study to show just how difficult it is to accurately estimate win probability using a statistical model.
Observational play-by-play data has a strong dependence structure that inflates the bias and variance of these estimators.
Further, to achieve approximately valid marginal coverage, win probability confidence intervals need to be substantially wide.
Concisely, these are high variance estimators subject to substantial uncertainty.

In future work, we suggest exploring probabilistic state-space models to estimate win probability (for real sports like American football).
Those models simplify the game of football into a series of transitions between game-states.
Transition probabilities are estimated from play-level data and win probability is calculated by simulating games.
The effective sample size ($\ESS$) is the number of plays because transition probabilities are fit from independent play-level observations.
Though state-space models have lower variance (via a higher $\ESS$) than statistical models, they have higher bias, as they make stronger simplifying assumptions.
% We look forward to a public facing exploration of those models in the future.

We also look forward to further research on the impact of a strong dependency structure in other sports applications.
The structure of play-by-play win probability data, in which groups of observations share the same outcome, is not unique.
It is prevalent across myriad sports datasets.
It appears in any dataset in which the outcome variable is the final result of some unit of time (e.g., a game or play) and the observations consist of units (e.g., frames or plays) leading to that end result.
For instance, expected points models are fit from play-by-play data for which large clusters of plays share the same next score outcome \citep{nflWar}.
This structure also appears in models fit from tracking data that map actions during each frame of a play to the ultimate outcome of a play.
For instance, \citet{YurkoMatanoRichardsonGraneredPospisilPelechrinisVentura+2020+163+182} use tracking data to model the expected yards gained for a ball-carrier during the course of a play. 
Each row (frame) within the same play shares the same outcome (yards gained on that play).
% In each of these other applications, we implore modelers to consider the ramifications of a strong dependency structure.

% %%%%%%%%%%%%%%%%%%%%%%%%%%%%%
% \bibliography{refs}
% \end{document}
% %%%%%%%%%%%%%%%%%%%%%%%%%%%%%

\newpage
\bibliography{refs}

% \clearpage
\begin{center}
{\large\bf SUPPLEMENTARY MATERIAL}
\end{center}
\appendix

%%%%%%%%%%%%%%%%%%%%%%%%%%%%%%%%%%%%%%%%%%%%%%%%%%%%%%%%%%%%%%%%%%%%%%%%%%%%%%%%%%%%%%
\section{Our code}

The code for this study is publicly available on GitHub at \url{https://github.com/snoopryan123/fourth_down} in the folder \texttt{1\_simulation/sim\_v3}.

\end{document}